\documentclass[aps,prb,twocolumn,superscriptaddress,showpacs]{revtex4}

\usepackage{graphicx, latexsym, verbatim}
\usepackage{amssymb, amsmath}
\usepackage[ansinew]{inputenc}
\usepackage{color}
\usepackage[english]{babel}

\begin{document}

\title{Improving Transition Voltage Spectroscopy of Molecular Junctions}

\author{Troels Markussen}
\affiliation{Center for Atomic-scale Materials Design (CAMD), Department of Physics,
Technical University of Denmark, DK-2800 Kgs. Lyngby, Denmark}
\affiliation{Danish National Research Foundations Center of Individual Nanoparticle Functionality (CINF), Department of Physics, Technical University of Denmark, DK-2800 Kgs. Lyngby, Denmark}

\author{Jingzhe Chen}
\email[]{jingzhe.chen@gmail.com}
\affiliation{Center for Atomic-scale Materials Design (CAMD), Department of Physics,
Technical University of Denmark, DK-2800
Kgs. Lyngby, Denmark}

\author{Kristian S. Thygesen}
\affiliation{Center for Atomic-scale Materials Design (CAMD), Department of Physics,
Technical University of Denmark, DK-2800 Kgs. Lyngby, Denmark}


\date{\today}

\begin{abstract}
Transition voltage spectroscopy (TVS) is a promising spectroscopic
tool for molecular junctions. The principles in TVS is to find the
minimum on a Fowler-Nordheim plot where $\ln(I/V^2)$ is plotted
against $1/V$ and relate the voltage at the minimum, $V_{\rm min}$,
to the closest molecular level. Importantly, $V_{\rm min}$, is
approximately half the voltage required to see a peak in the $dI/dV$
curve. Information about the molecular level position can thus be
obtained at relatively low voltages. In this work we show that the
molecular level position can be determined at even lower voltages,
$V_{\rm min}^{(\alpha)}$ by finding the minimum of $\ln(I/V^\alpha)$
with $\alpha<2$. On the basis of a simple Lorentzian transmission
model we analyze theoretical {\it ab initio} as well as experimental
$I-V$ curves and show that the voltage required to determine the
molecular levels can be reduced by $\sim 30\%$ as compared to
conventional TVS. As for conventional TVS, the symmetry/asymmetry of
the molecular junction needs to be taken into account in order to
gain quantitative information. We show that the degree of asymmetry
may be estimated from a plot of $V_{\rm min}^{(\alpha)}$ vs.
$\alpha$.
\end{abstract}


\maketitle
\section{Introduction}
The field of molecular electronics holds the promise of low-cost and
flexible devices and potentially a continued miniaturization of
electronic circuits beyond the limits of standard silicon
technologies.\cite{poulsen09,SongNature2009} The electronic
properties of a molecular junction are to large extend governed by
the position of the frontier orbitals with respect to the electrode
Fermi levels, i.e. either the highest occupied molecular orbital
(HOMO) or the lowest unoccupied molecular orbital (LUMO). One way of
measuring the energetic position of the HOMO (or LUMO) level is from
peaks in the $\text{d}I/\text{d}V$ curve, where $I$ is the current
and $V$ is the bias voltage. In the simplest picture, it requires a
voltage of $V \sim 2|E_F-\varepsilon_H|/e$ to probe the HOMO
position in a symmetric junction. However, in practice the junction
will often suffer from breakdown before the peak in
$\text{d}I/\text{d}V$ is reached.

Transition voltage spectroscopy (TVS) has recently been introduced as a spectroscopic tool in molecular electronics~\cite{Beebe2006} and offers the possibility to probe the energy of molecular levels at relatively low voltages. Beebe \textit{et al.}~\cite{Beebe2006} showed that the HOMO position, determined from ultraviolet photo spectroscopy (UPS), scaled linearly with a characteristic voltage, $V_{\rm min}$ obtained at the minimum of a Fowler-Nordheim graph, i.e. a plot of $\ln(I/V^2)$ against $1/V$. TVS is now becoming an increasingly popular tool in molecular electronics~\cite{Noguchi2007,Beebe2008,Zangmeister2008,HoChoi2008,SongNature2009,TanAPL2010,Smaali2010}. The great advantage of TVS is that the minimum voltage $V_{\rm min}\approx |E_F-\varepsilon_H|$ in a symmetric junction is obtained at approximately half the voltage required for measuring a peak in $\text{d}I/\text{d}V$.\cite{AraidaiPRB2010,ChenMarkussenPRB2010}

The original interpretation of TVS is based on a tunnel barrier model~\cite{Simmons}. Within this model the transition voltage is reached when the tunnel barrier, due to the bias voltage, is shifted from a trapezoidal to a triangular shape. However, Huismann {\it et al.}~\cite{Huisman2009} pointed out that the barrier model in some respects is inconsistent with experiments and advocated that a Landauer transmission model with a single electronic level is more appropriate. In the Landauer model, the transmission function is assumed to be described by a single Lorentzian with a peak at the HOMO position. In a recent work~\cite{ChenMarkussenPRB2010} we studied a large number of molecular contacts by {\it ab initio} transport calculations, and showed that the simple Lorentzian transmission is a proper model, provided that the symmetry/asymmetry of the junction is taken into account.

When a bias voltage is applied across a molecular junction, the two Fermi levels of the electrodes open a bias window in the gap between the HOMO and the LUMO. The $dI/dV$ curve is peaked, when one of the Fermi levels is aligned with either the HOMO or the LUMO.
At lower voltages, the Fermi levels go through the tail of the corresponding Lorentzian peak. The shape of the tail is determined by the level position, $\varepsilon_0$, and the broadening, $\Gamma$, which is influenced by the contact geometries. As illustrated in Ref.~\onlinecite{ChenMarkussenPRB2010} the TVS minimum voltage is insensitive to the broadening provided that $|\varepsilon_0-E_F|/\Gamma\gg1$.

Our main focus in this paper is to discuss how to measure the HOMO or LUMO position at low voltages by exploiting characteristic features in the tail of the Lorentzian transmission peak. To this end we extend the TVS to a more general form, i.e., the objective function is switched to $\ln(I/V^\alpha)$ with $\alpha\leq 2$, where we recover the usual TVS for $\alpha=2$. The minimum of $\ln(I/V^\alpha)$ is denoted $V_{\rm min}^{(\alpha)}$. The properties of this function include: (i) the linear relationship between $V_{\rm min}^{(\alpha)}$ and $\varepsilon_0$ still exists which means any $\alpha$ value can in principle be used to predict $\varepsilon_0$; (ii), a minimum value in the general TVS plot can always be obtained by tuning the $\alpha$ value; (iii), compared with conventional TVS, $V_{\rm min}$  can be even lower with smaller $\alpha$. The last point is a substantial technical improvement since the importance of TVS lies in the fact that the minimum in the Fowler-Nordheim plot is obtained at relatively low bias voltage. The analysis presented in this article
based on theoretical {\it ab initio} as well as experimental $I-V$-curves shows that the voltage required to determine the molecular levels can be reduced by $\sim 30\%$. As previously mentioned, the symmetry/asymmetry of the molecular junction needs to be taken into account in order to gain quantitative information. We show that the degree of asymmetry may be estimated from a plot of $V_{\rm min}^{(\alpha)}$ vs. $\alpha$.

\section{Model}
As introduced in Ref.~\onlinecite{ChenMarkussenPRB2010},  the energy-dependent transmission coefficient is assumed to be a Lorentzian
\begin{equation}
 T(E;\varepsilon_0,\Gamma) = \frac{f}{(E-\varepsilon_0)^2+\Gamma^2/4},
\label{Transmission}
\end{equation}
where $\varepsilon_0$ and $\Gamma$ are the molecular level energy and broadening,
respectively. $f$ is a constant factor to account for multiple molecules in the
junction, asymmetric coupling to the electrodes, etc. Our analysis will not be
dependent on the actual value of $f$. We shall assume that the molecular level is the HOMO level, i.e. $\varepsilon_0<E_F$. This is the typical case for thiol bonded molecules (as considered in this work), where electron transfer from the metal to the sulfur end group shifts the molecular levels upward in energy\cite{guowen09}.
At finite bias voltage, the molecular level may be shifted relative to the zero-bias position. Such non-linear effects express themselves differently in symmetric and asymmetric junctions. We describe the degree of asymmetry by the parameter $\eta\in[-1/2; 1/2]$ such that the current is given by
\begin{equation}
I = \frac{2e}{h}\int_{-\infty}^{\infty}T(E;\varepsilon_0+\eta
\,V,\Gamma)\left[f_L(V/2)-f_R(-V/2) \right]{\rm d}E, \label{Current}
\end{equation}
where $f_{L/R}(V)=1/[\exp{(E_F+ e V)/k_B T}+1]$ are the Fermi-Dirac distributions for the left and right contact, respectively. A symmetric junction has $\eta=0$, while a completely asymmetric junction has $\eta=\pm 1/2$.

\begin{figure}[htb!]
\includegraphics[width=0.85\columnwidth]{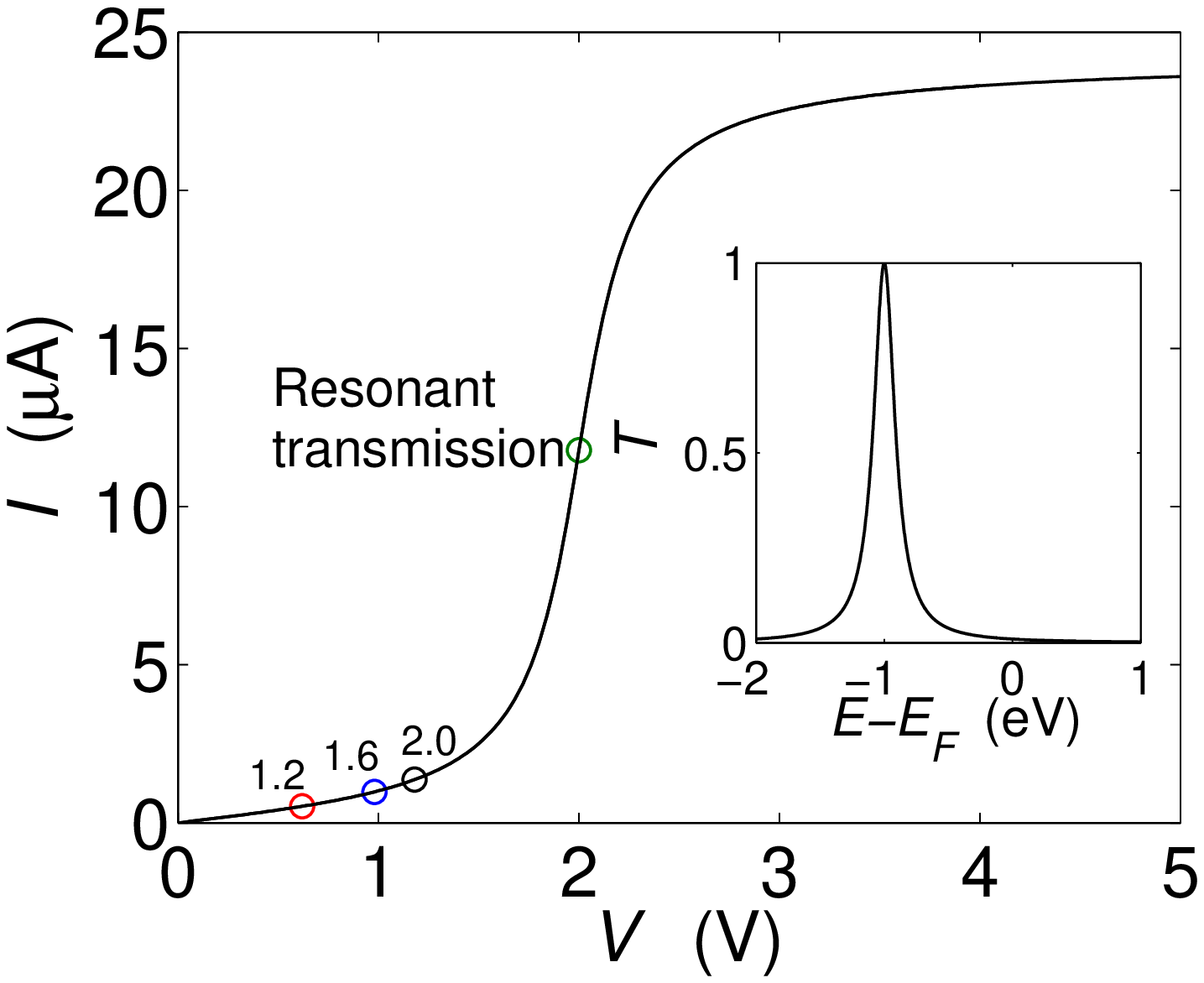}
\includegraphics[width=0.85\columnwidth]{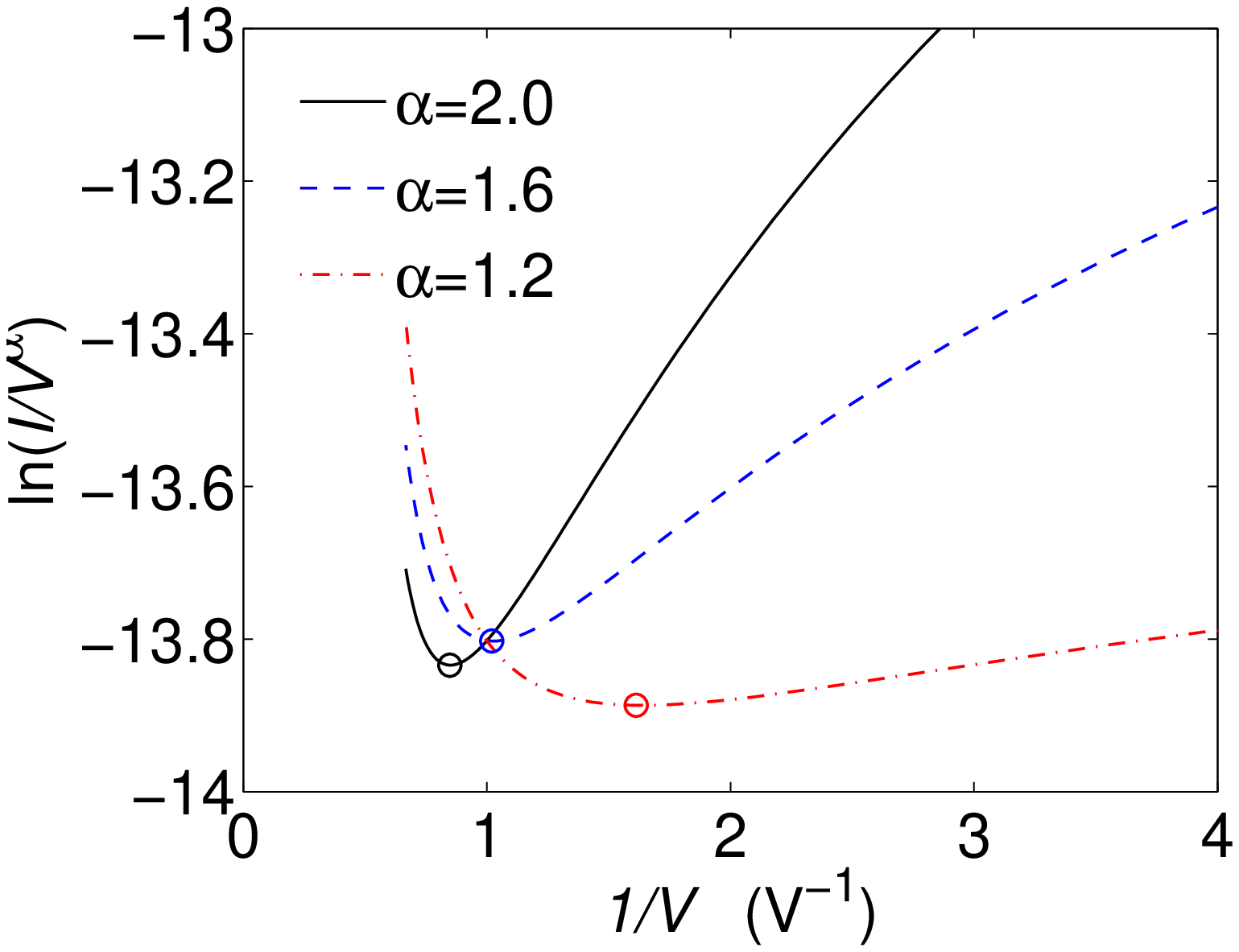}
    \caption{Top: $I-V$ curve from a symmetric junction and a Lorentzian transmission function with $\varepsilon_0-E_F=-1\,$eV and $\Gamma=0.2\,$eV (shown in the inset). The three lower points on the curve mark the voltages, $V_{\rm min}^{(\alpha)}$ at the the minima of Fowler-Nordheim plots $\ln(I/V^\alpha)$ vs. $1/V$. The numbers show the $\alpha$ values. The resonant transmission voltage is also marked. Bottom: Generalized Fowler-Nordheim plots where $\ln(I/V^\alpha)$ is plotted against $1/V$. }
\label{IV_model}
\end{figure}

Figure \ref{IV_model} (top) shows the $I-V$ curve for a symmetric
junction with a Lorentzian transmission function with
$\varepsilon_0-E_F=-1\,$eV and $\Gamma=0.2\,$eV. The points marked
by circles represent the minima of the $\ln(I/V^\alpha)$ vs. $1/V$
plots shown in the bottom panel. The resonant transmission happens
at the bias voltage $V=2\,$V, which can be reduced to $1.2\,$V by
using the conventional TVS ($\alpha=2.0$). Using a smaller $\alpha$
value, the minimum can be obtained at even lower bias, i.e.,
$1.0\,$V and $0.6\,$V for $\alpha=1.6$ and $\alpha=1.2$
respectively. We thus see that by using $\alpha=1.2$, the voltage
needed to measure the HOMO position is reduced by a factor of three
compared to the resonant transmission and is only half the voltage
required using conventional TVS with $\alpha=2$. However, it is also
evident from Fig. \ref{IV_model} (bottom) that the minima in the TVS
plots for small $\alpha$ values are more shallow and it might be
difficult to accurately determine the minimum if the $I-V$ data
contains noise. Consequently, the uncertainty of TVS increases with
decreasing $\alpha$.

\begin{figure}[htb!]
\includegraphics[width=0.85\columnwidth]{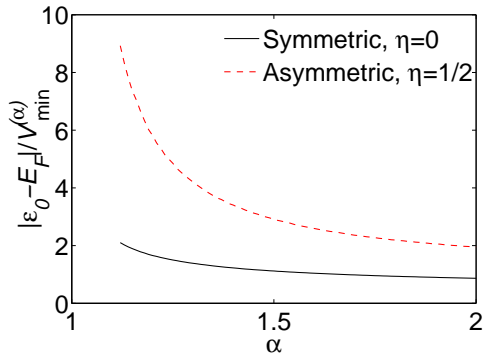}
    \caption{Ratio of the resonant level and minimum voltage, $|\varepsilon_0-E_F|/V_{\rm min}^{(\alpha)}$ vs. $\alpha$ obtained for a symmetric ($\eta=0$) and completely asymmetric ($\eta=1/2$) junction.}
\label{ratio_vs_alpha}
\end{figure}

We conclude the analysis of the Lorentzian transmission model by considering the effects of asymmetry. Figure \ref{ratio_vs_alpha} shows the \textit{ratio} of the resonant level and minimum voltage vs. $\alpha$ obtained for a symmetric and completely asymmetric junction. For both junctions, the ratio increases by decreasing the $\alpha$ value. In accordance with Fig. \ref{IV_model} this implies that for a given value of $\varepsilon_0$, the generalized TVS minimum, $V_{\rm min}^{(\alpha)}$ is \textit{reduced} by decreasing $\alpha$. From Fig. \ref{ratio_vs_alpha} it is seen that the reduction in $V_{\rm min}^{(\alpha)}$ is the largest for a completely asymmetric junction ($\eta=1/2$), where the minimum voltage is reduced by a factor of 4 in tuning $\alpha$ from 2 to 1.2. In the same $\alpha$-range, the minimum for a symmetric junction is reduced by a factor of 2.



\section{{\it Ab initio} calculations}
Having established that the generalized TVS plot of $\ln(I/V^\alpha)$ can provide information of the molecular level position at significantly lower voltages than required by conventional TVS (using $\alpha=2$), we now turn to analyze more realistic $I-V$ curves calculated from first principles.


The current at finite bias voltage is calculated using DFT in combination with a non-equilibrium Green function (NEGF) method as described in Ref. \onlinecite{gpaw-review}.  Our DFT-NEGF method is implemented in GPAW, which is a real space electronic structure code based on the projector augmented wave method~\cite{gpaw,gpaw-lcao}. We use the Perdew-Burke-Ernzerhof (PBE) exchange-correlation functional~\cite{PBE}, and a $4\times4$ $k$-point sampling in the surface plane. The electronic wave functions are expanded in an atomic orbital basis~\cite{gpaw-lcao}. In all calculations, the molecule and the closest Au layers are described by a double-zeta plus polarization (dzp) basis set, while the remaining Au atoms are described by a single-zeta plus polarization (szp) basis.

\begin{figure}[htb!]
%
%
\includegraphics[width=\columnwidth]{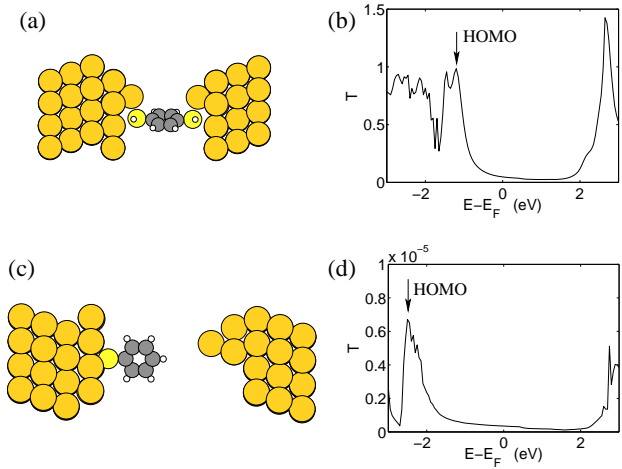}
    \caption{Atomic structure of a symmetric Au-BDT-Au junction (a) together with an asymmetric Au-Ph-STM junction (c). The corresponding transmission functions are shown in panel (b) and (d), respectively.}
\label{structures}
\end{figure}

First, we investigate two proto-typical molecular junctions, Au-1,4-benzenedithiol-Au (Au-BDT-Au) and Au-phenyl-STM (Au-Ph-STM) as shown in  Fig.~\ref{structures} (a) and (c), representing the symmetric and asymmetric class respectively. From the transmission plot in Fig.~\ref{structures} (b) we can see for the Au-BDT-Au junction, the HOMO level is about $-1.2\,$eV and LUMO level is about $2.8\,$eV. Since the bias window opens symmetrically, the TVS measurement will be a measurement of the HOMO position.  The transmission for the Au-Ph-STM junction, Fig. \ref{structures} (d), shows that the HOMO level is about $-2.6\,$eV and LUMO level is about $2.8\,$eV. Since it is a highly asymmetric junction, the HOMO level will be more pinned to the Fermi level of left electrode, and when a positive bias voltage is applied the Fermi level of the right electrode will scan through the HOMO. The TVS plot will thus again be a measurement of the HOMO position. A detailed analysis of the transmission spectra~\cite{ChenMarkussenPRB2010} at finite bias voltages shows that the junction is not completely asymmetric, but rather has $\eta=0.33$. This is due to the relative small vacuum gap between the right tip and the molecule causing a finite voltage drop over the molecule itself. Increasing the vacuum gap further would increase $\eta$ but this would cause computational problems due to the finite range of the atomic orbital basis.

\begin{figure}[htb!]
\includegraphics[width=0.8\columnwidth]{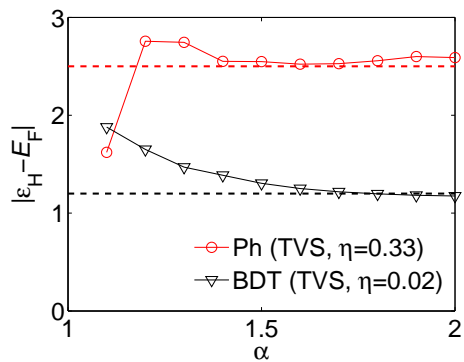}
    \caption{Predicted HOMO level of phenyl (Ph) and benzenedithiol (BDT) junctions in Fig. \ref{structures} obtained from the TVS spectra using different $\alpha$-values. The horizontal dashed lines mark the HOMO position as determined from the transmission spectra in Fig. \ref{structures} (b) and (d).}
\label{predicted_Eh}
\end{figure}

From the calculated $I-V$ curves (not shown) we find the generalized TVS minimum voltages, $V_{min}^{(\alpha)}$, for different $\alpha$ values. Based on curves like the ones shown in Fig. \ref{ratio_vs_alpha} for $\eta=0.04$ and $\eta=0.33$ we obtain {\it predicted} values of the HOMO position at different $\alpha$ values. These predicted HOMO values are shown in Fig. \ref{predicted_Eh} vs. $\alpha$ and they are seen to converge to the ``true'' HOMO position as determined from the transmission plots (Fig. \ref{structures} (b) and (d)) as $\alpha$ is increased. For both junction, using $\alpha=1.5$ is sufficient to predict the HOMO position with an error around $0.1\,$eV. This corresponds to a reduction in TVS minimum voltages by 23\% and 31\% for the symmetric (Au-BDT-Au) and asymmetric (Au-Ph-STM) junctions, respectively.

It is well known that DFT is unable to accurately describe energy gaps and level alignment of molecules at surfaces~\cite{NeatonPRL2006,juanma09}. The HOMO positions calculated in Fig. \ref{predicted_Eh} can therefore not be expected to be in quantitative agreement with the experimental value. However, it is not our goal in this paper to use DFT to predict the HOMO position, but rather to show that the generalized TVS scheme can be applied on realisticly shaped $I-V$ curves beyond simple models. In this respect, and since the ratio $|\varepsilon_0-E_F|/V^{(\alpha)}_{\rm min}$ is independent on $\varepsilon_0$, the results in Fig. \ref{predicted_Eh} are still valid, irrispectively of the inaccuracies in DFT.

\begin{figure}[htb!]
\includegraphics[width=0.9\columnwidth]{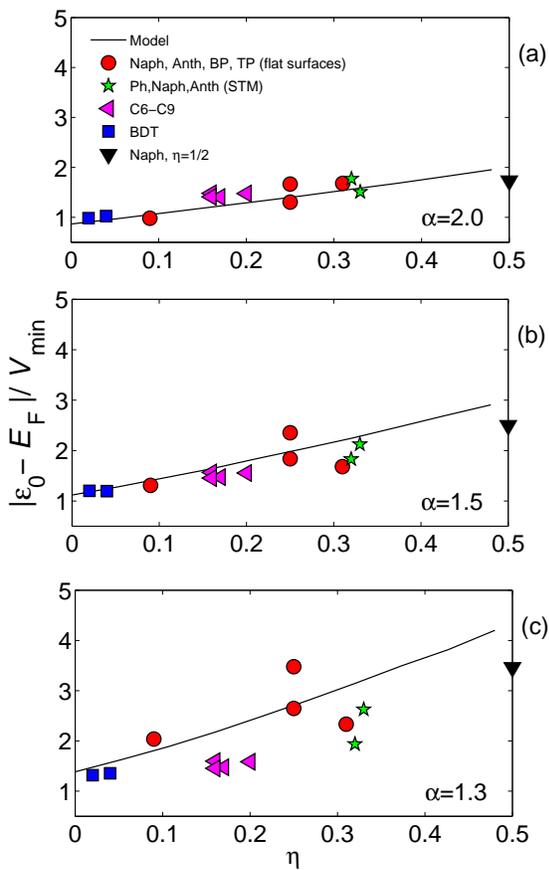} 
    \caption{Ratio of HOMO position and minimum voltage, $|\varepsilon_H-E_F|/V_{\rm min}^{(\alpha)}$ vs. asymmetry parameter $\eta$. The minimum voltages are obtained with the generalized TVS scheme using $\alpha=2.0$ (a), $\alpha=1.5$ (b), and $\alpha=1.3$ (c). The {\it ab initio} calculations follow the Lorentzian model prediction for $\alpha=1.5$ and $\alpha=2.0$, while substantial deviations are seen for $\alpha=1.3$.}
\label{e0Vmin_vs_eta}
\end{figure}

To get a broader overview of the feasibility of the generalized TVS we analyze a series of molecular junctions. All the original data we use here are the same as those in our previous paper~\cite{ChenMarkussenPRB2010}. Figure \ref{e0Vmin_vs_eta} shows the ratio $|\varepsilon_H-E_F|/V_{\rm min}^{(\alpha)}$ vs. asymmetry parameter $\eta$. As also shown in Ref. \onlinecite{ChenMarkussenPRB2010}, the {\it ab initio} data for $\alpha=2$ (panel a) agree well with the analytical model when the asymmetry is taken into account. Importantly, a similar good agreement between the DFT data and the model is seen for $\alpha=1.5$ (panel b). It thus seems to be of general validity that quantitative information can be obtained using $\alpha=1.5$. Further decreasing $\alpha$ to 1.3 (panel c) leads to significantly larger deviations from the model calculation.  The deviation can be due to (i) the transmission peak differs from a Lorentzian; (ii) error in the asymmetry factor measurement because of non-linear effects; and (iii) for small values of $\alpha$ the determination of the TVS minimum voltage becomes uncertain. 

For practical purposes, it may be usefull to apply the approximate linear relations between  $|\varepsilon_H-E_F|/V_{\rm min}^{(\alpha)}$ and $\eta$:
\begin{eqnarray}
|\varepsilon_H-E_F|/V_{\rm min}^{(\alpha)}&=&2.3\,\eta + 0.85\;\;\;\;(\alpha=2) \label{linear-alpha-2} \\
|\varepsilon_H-E_F|/V_{\rm min}^{(\alpha)}&=&3.7\,\eta + 1.1\;\;\;\;(\alpha=1.5) \label{linear-alpha-1.5}.
\end{eqnarray}
If knowledge of the asymmetry parameter exists, the linear relations can be used to predict the HOMO (or LUMO) position using the general TVS scheme. We note in particular, that if the same predicted HOMO level is obtained from both Eq. \eqref{linear-alpha-2} and \eqref{linear-alpha-1.5} this is a strong indication that the true transmission function is well described by a Lorentzian function and the above theory is applicable. On the contrary, if information about the HOMO position exist, e.g. from UPS experiments~\cite{Beebe2006}, Eqs. \eqref{linear-alpha-2}-\eqref{linear-alpha-1.5} can be used to determine the junction asymmetry.

Based on the {\it ab initio} calculations, we conclude that
$\alpha=1.5$ can be used in the general TVS measurement to obtain
the HOMO (or LUMO) position without losing precision. Depending on
the degree of junction asymmetry this reduces the required bias
range with 20-30\% as compared to conventional TVS.


\section{Analysis of experimental data}
\begin{figure}[htb!]
\includegraphics[width=0.9\columnwidth]{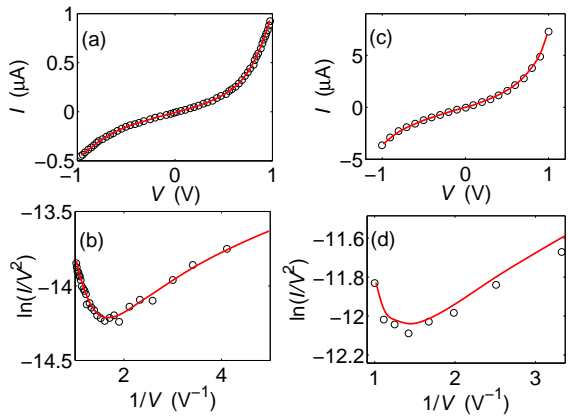}
    \caption{(a): $I-V$ curve acquired from Ref.~\onlinecite{Beebe2006} (circles) together with a smooth curve obtained by convolution of the data points with a Gaussian function. (b) Fowler-Nordheim plot (using $\alpha=2$) for the data points and the smooth function. (c) and (d) show similar data obtained from Ref~\onlinecite{TanAPL2010}.}
\label{data_Beebe}
\end{figure}

In the previous section it was shown that the generalized TVS based on a Lorentzian transmission function provides a good description of {\it ab initio} calculations. In this section we proceed to analyze experimental $I-V$ data using the generalized TVS. We obtain the experimental data from Beebe {\it et al.}~\cite{Beebe2006} and Tan {\it et al.}~\cite{TanAPL2010}. We manually read off the experimental data, and subsequently smoothen the data by a Gaussion convolution. In Fig. \ref{data_Beebe} we show the obtained $I-V$ data in panel (a) for a Au-anthracenethiol-Au junction~\cite{Beebe2006}  and  in panel (c) for a Au-terphenyl-thiol-Au junction~\cite{TanAPL2010}. Both experiments are performed with mono-thiol molecules using conducting atomic force microscope (AFM). The solid lines show the data after the smoothening. Panels (b) and (d) show the corresponding Fowler-Nordheim plot using $\alpha=2$.

As shown in Ref. \onlinecite{ChenMarkussenPRB2010} and discussed above, quantitative use of TVS is only possible if knowledge about the junction asymmetry exist. However, it may be difficult to measure this quantity in experiments directly. While it is certainly evident that the $I-V$ curves in Fig. \ref{data_Beebe} (a) and (c) are  asymmetric, it is difficult to get a quantitative estimate of the asymmetry factor from these values alone.

\begin{figure}[htb!]
\includegraphics[width=0.8\columnwidth]{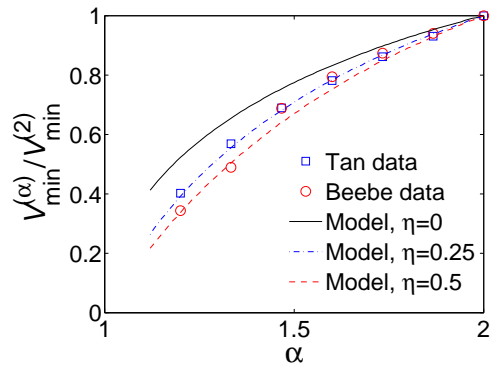}
    \caption{Ratio of $V_{\rm min}^{(\alpha)}/V_{\rm min}^{(2)}$ obtained from the experimental data (symbols) and from the Lorentzian model (lines). The experimental data follow the model reasonably well for $\eta=0.25$. This asymmetry value is in good agreement with our previous calculations~\cite{ChenMarkussenPRB2010}.}.
\label{Vmin_alpha_Beebe_Tan}
\end{figure}

Within the Lorentzian transmission model $V_{\rm min}^{(\alpha)}$ is
proportional to $|\varepsilon_0-E_F|$ provided that
$|\varepsilon_0-E_F|\gg\Gamma$. The proportionality constant depends
on $\alpha$ and on the asymmetry factor, $\eta$. This implies that
the {\it ratio}  $V_{\rm min}^{(\alpha)}/V_{\rm min}^{(2)}$ is
independent of both $|\varepsilon_0-E_F|$ and $\Gamma$. We may thus
in principle determine $\eta$ from the dependence of $V_{\rm
min}^{(\alpha)}/V_{\rm min}^{(2)}$ as function of $\alpha$. In Fig.
\ref{Vmin_alpha_Beebe_Tan} we plot $V_{\rm min}^{(\alpha)}/V_{\rm
min}^{(2)}$ vs. $\alpha$ for the two experimental data sets together
with the model calculations based on a Lorentzian transmission,
corresponding to three different asymmetry parameters. We first
notice that the experimental data points fall within the model
calculations corresponding to a completely symmetric ($\eta=0$)
junction and a completely asymmetric junction ($\eta=0.5$). This
indicates that the true transmission functions indeed is well
described by a Lorentzian shape. Secondly, we also see that both
experimental data sets follow approximately the model calculation
corresponding to  $\eta=0.25$. We note that this asymmetry value is
in good agreement with our previous
calculations~\cite{ChenMarkussenPRB2010} and we will use $\eta=0.25$
in the remaining analysis.

\begin{figure}[htb!]
\includegraphics[width=0.8\columnwidth]{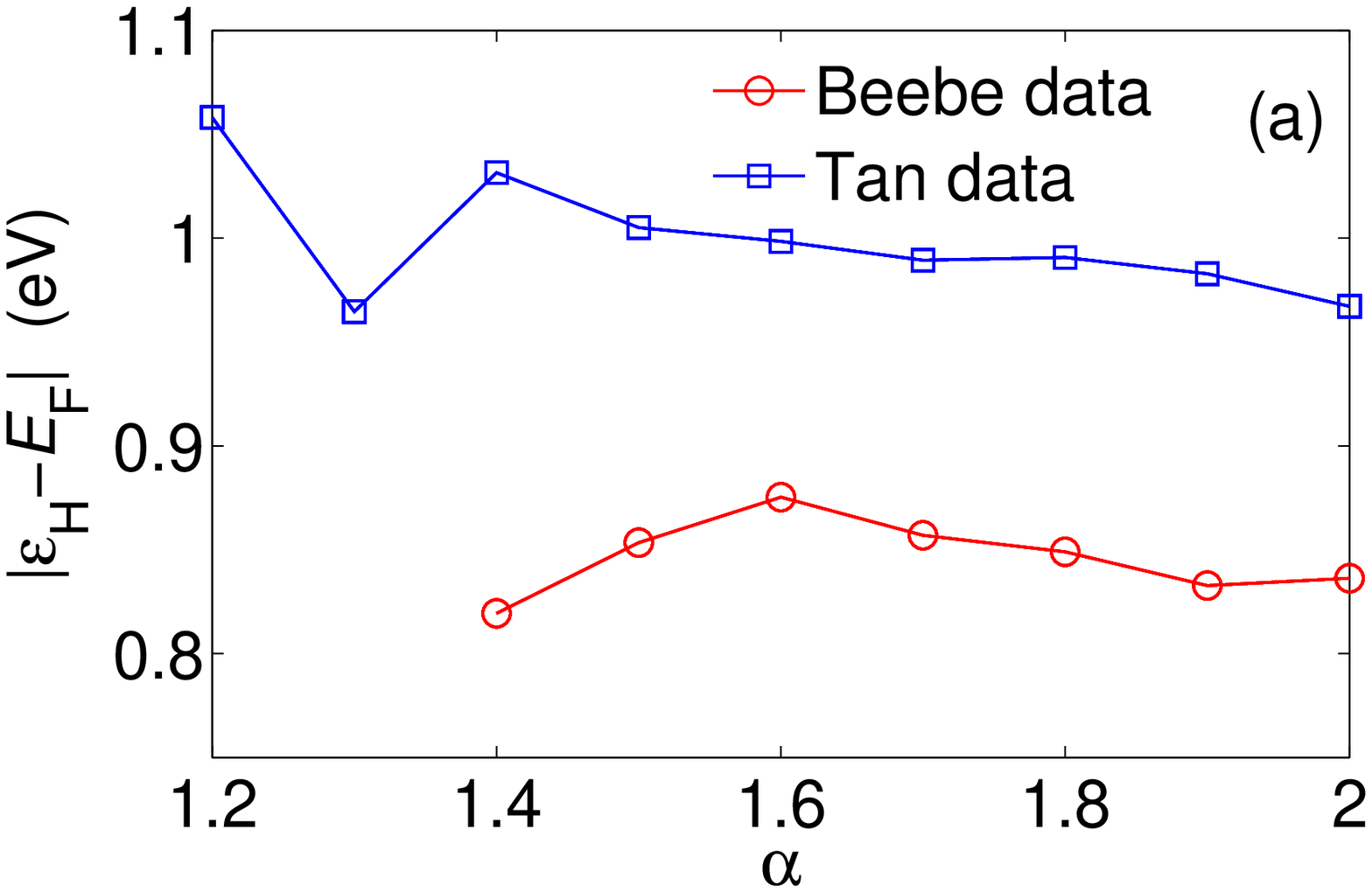}

\includegraphics[width=0.8\columnwidth]{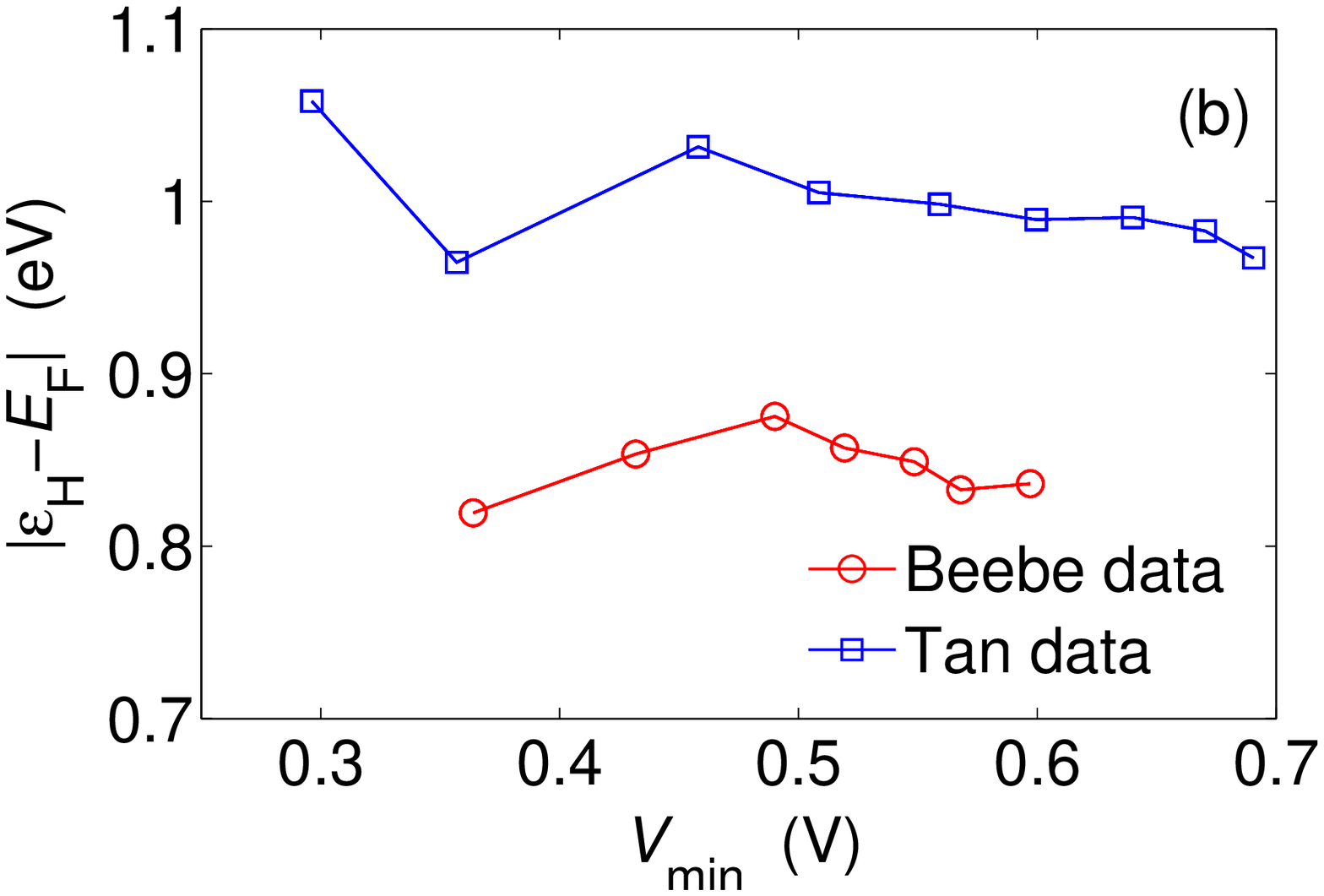}
    \caption{Predicted values of $|\varepsilon_H-E_F|$ vs. $\alpha$ (a) and vs. $V_{\rm min}$ (b) for anthracene (Beebe data) and terphenyl (Tan data) junctions. The predicted HOMO levels are reasonably constant for $\alpha>1.5$, thus indicating that the Lorentzian model is applicable in analyzing the experimental data. Panel (b) shows that using $\alpha\approx1.5$, the HOMO level can be determined at $\sim$30\% lower voltage than required for $\alpha=2$.}
\label{predicted_HOMO_Beebe_Tan}
\end{figure}

From the generalized TVS minimum voltages $V_{\rm min}^{(\alpha)}$ obtained from the experimental data together with a model calculations like in Fig. \ref{ratio_vs_alpha} (for $\eta=0.25$) we obtain {\it predicted} HOMO values. The results are shown in Fig. \ref{predicted_HOMO_Beebe_Tan} vs. $\alpha$ in panel (a) and vs.  minimum voltage in panel (b). As previously  seen for the {\it ab initio} calculations, the predicted HOMO values are reasonably well converged for $\alpha\geq 1.5$ for both data sets. As seen in panel (b) this corresponds to minimum voltages reduced by $\sim$30\%.

From Fig. \ref{predicted_HOMO_Beebe_Tan} we obtain approximate predicted values of $|\varepsilon_H-E_F|\approx0.85\,$eV and 1.0 eV for Beebe (anthracene) and Tan (terphenyl) data, respectively.  The HOMO positions of anthracene-thiol and terphenyl-thiol on a Au surface were determined from ultraviolet photoelectron spectroscopy (UPS) to be $E_F-\varepsilon_H= 1.6\,$eV and $E_F-\varepsilon_H= 1.7\,$eV, respectively.\cite{Beebe2006}  The presence of the AFM tip in the transport
measurements will lead to a renormalization of the HOMO energy due to an image charge effect~\cite{juanma09} which is not present in the UPS measurements. Based on the method described in Ref. \onlinecite{duncan} we estimate the image charge interaction to be $\sim0.5\,$eV. Correcting the UPS values for $\varepsilon_H$ by this value ($\tilde{\varepsilon}_H=\varepsilon_H+0.5\,$eV) lead to  $E_F-\tilde{\varepsilon}_H \approx 1.1\,$eV for the anthracene and $E_F-\tilde{\varepsilon}_H \approx 1.2\,$eV for terphenyl, which agree with our TVS predicted HOMO values within 0.25 eV, corresponding to a deviation of $\sim25\,$\%. The original interpretations of TVS assuming a one-to-one relation between the HOMO position and the TVS minimum voltage (i.e. $|\varepsilon_H-E_F|=V_{\rm min}^{(\alpha=2)}$) would have resulted in values of $|\varepsilon_H-E_F|=0.62,\;0.69\;$eV. These values differs significantly from the image charge corrected UPS values by $\sim 75\,$\%. The closer agreement between our predicted HOMO values and the UPS measurements signifies the importance of taking the asymmetry of the molecular junction into account.

\section{Summary}
Transition voltage spectroscopy (TVS) is a valuable spectroscopic tool for molecular junctions as it offers the opportunity to probe molecular levels at relatively low voltages. In this work we have generalized the theory of TVS by showing that the minimum of $\ln(I/V^\alpha)$, with $\alpha<2$, provides quantitative information of the HOMO position at {\it lower} bias voltages than required by conventional TVS, using $\alpha=2$. On the basis of a simple Lorentzian transmission model we have analyzed theoretical {\it ab initio} as well as experimental $I-V$-curves and show that the voltage required to determine the molecular levels can be reduced by $\sim 30\%$ as compared to conventional TVS. Importantly, the symmetry/asymmetry of the molecular junction needs to be taken into account in order to gain quantitative information. We believe the present work will be valuable for future theoretical as well as experimental work in the characterization of molecular junctions.

%
%

\begin{acknowledgements}
The center for Atomic-scale Materials Design (CAMD) is funded by the
Lundbeck Foundation. The Center for Individual Nanoparticle
Functionality (CINF) is sponsored by the Danish National Research
Foundation. The authors acknowledge support from FTP through grant
no. 274-08-0408 and from The Danish Center for Scientific Computing.
\end{acknowledgements}


\end{document}